\documentclass[12pt]{article}
\usepackage{graphics}
\def\al{\alpha}
\def\be{\beta}
\def\ga{\gamma}
\def\de{\delta}
\def\ep{\epsilon}

\def\th{\theta}

\def\si{\sigma}

\def\om{\omega}

\def\phit{{\tilde\varphi}}

\def\De{\Delta}

\def\half{{1\over{2}}}
\def\pd{\partial}

\newcommand{\beq}{\begin{equation}}
\newcommand{\eeq}{\end{equation}}
\newcommand{\bea}{\begin{eqnarray*}}
\newcommand{\eea}{\end{eqnarray*}}
\newcommand{\beaq}{\begin{eqnarray}}
\newcommand{\eeaq}{\end{eqnarray}}
\begin{document}
\begin{flushright}APCTP-1999007\\KIAS-P99018\end{flushright}
\centerline{\Large \bf Hidden Relation between Reflection Amplitudes}
\centerline{\Large \bf and Thermodynamic Bethe Ansatz}
\vskip 1cm
\centerline{\large Changrim Ahn\footnote{ahn@mm.ewha.ac.kr},
Chanju Kim\footnote{cjkim@ns.kias.re.kr},
and Chaiho Rim\footnote{rim@pine.chonbuk.ac.kr}}
\vskip 1cm
\centerline{\it$^{1}$Department of Physics, Ewha Womans University}
\centerline{\it Seoul 120-750, Korea}
\vskip .5cm
\centerline{\it $^{2}$ School of Physics, Korea Institute for Advanced Study}
\centerline{\it Seoul, 130-012, Korea}
\vskip .5cm
\centerline{\it $^{3}$ Department of Physics, Chonbuk National University}
\centerline{\it Chonju 561-756, Korea}
\vskip 1cm
\centerline{\small PACS: 11.25.Hf, 11.55.Ds}
\vskip 2cm
\centerline{\bf Abstract}

In this paper we compute the scaling functions of the effective central 
charges for various quantum integrable models in a deep ultraviolet
region $R\to 0$ using two independent methods.
One is based on the ``reflection amplitudes'' of the (super-)Liouville
field theory where the scaling functions are given by the 
conjugate momentum to the zero-modes.
The conjugate momentum is quantized for the sinh-Gordon, the Bullough-Dodd, 
and the super sinh-Gordon models where the quantization conditions
depend on the size $R$ of the system and the reflection amplitudes.
The other method is to solve the standard thermodynamic Bethe ansatz (TBA) 
equations for the integrable models
in a perturbative series of $1/({\rm const.}-\ln R)$. 
The constant factor which is not fixed in the lowest order computations
can be identified {\it only when} we compare the higher order corrections
with the quantization conditions.
Numerical TBA analysis shows a perfect match for the scaling functions
obtained by the first method.
Our results show that these two methods are complementary to each other.
While the reflection amplitudes are confirmed by the numerical TBA analysis,
the analytic structures of the TBA equations become clear only when the
reflection amplitudes are introduced.

\newpage

\section{Introduction}

In the study of the two-dimensional quantum field theories near critical
points, perturbed conformal field theory (CFT) approach has been quite
successful \cite{sasha}. 
Certain perturbations maintain the integrability structures for the models
so that one can use the exact $S$-matrices and the thermodynamic 
Bethe ansatz (TBA) methods to compute various physical quantities,
in particular, the scaling function of the effective central 
charge as a function of the size of a system $R$ \cite{alyosha}.
In the vicinity of the ultraviolet (UV) fixed point,
the scaling functions of various models behave in different ways.
While most common behaviour is the power law corrections of $R$,
slow flows due to the $1/(a-\ln R)^2$ corrections have been found
for various affine Toda field theories with an unknown constant $a$.
We show in this paper that to determine the UV behaviour completely
one needs an independent method to compute the scaling function. 

The independent method we need has been first constructed in a remarkable 
paper by A. Zamolodchikov and Al. Zamolodchikov \cite{ZamZam}
where they introduce the ``reflection amplitude'' 
of the Liouville field theory (LFT) in terms of the correlation 
functions of the exponential operators and their dual fields.
With the reflection amplitudes explicitly constructed from the
structure constants of the LFT,
they derived the scaling function of the ShG model as
a function of a momentum which is conjugate to the bosonic zero-mode.
Considered as an integrable perturbation of the LFT, the ShG model 
provides a confining potential well for the zero-mode so that the
conjugate momentum should satisfy certain ``quantization condition''
which relates the momentum with $R$ depending on the details of the 
reflection amplitudes.
Numerical analysis shows a perfect agreement between the two scaling
functions, confirming the new method.

Our objectives in this paper is to compare the two methods both
analytically and numerically in a deep UV region for the 
Bullough-Dodd (BD) model, another integrable perturbation of
the LFT and to extend whole formalism to the supersymmetric case.
Formalism to study TBA equations analytically in the UV region
has been first presented in \cite{roaming} for the ShG model
by changing the nonlinear integral equation into infinite order
differential equation and has been extended to affine Toda theories
in \cite{martins,fring}.
However, these works have considered only leading corrections
of order $1/(a-\ln R)^2$. 
According to our analysis, the real interesting feature arises
only when one takes into account higher order corrections
where the quantization conditions found in the reflection amplitudes 
approach appear in the analysis of the TBA equations
as a hidden structure.

Comparison of these two methods can be used as a tool to check 
nonperturbative relations between masses of on-shell particles
and dimensional parameters appearing in the actions.
This is because the reflection amplitudes defined as off-shell 
quantities depend on the dimensional parameters while the TBA 
concerns only the on-shell quantities like the particle masses.
The relation for the ShG model in \cite{zammass} and generalizations
to (fractionally) supersymmetric theories in \cite{ssgmass} can be
tested against the TBA equations.

Another interesting point happens when we analyze the supersymmetric 
sinh-Gordon (SShG) model using the two methods. 
As a perturbed super-LFT, there are two reflection amplitudes 
corresponding to the Neveu-Schwarz (NS) and Ramond (R) sectors, 
respectively.
Being different, they generate different scaling behaviours.
Interestingly, the scaling function for the (R) sector has at most
the power law corrections only. 
While we have well-defined TBA equation for the (NS) sector, 
that for the (R) sector is not estabilished.
We suggest the TBA of the (R) sector and provide the justifications
based on the behaviour of the reflection amplitude.

This paper is organized as follows.
We introduce in sect. 2 the reflection amplitudes
for the LFT and super-LFT and
show that one can interpret the amplitudes as the quantum
mechanical reflection amplitudes of the wave function.
As an independent check, we solve the zero-mode quantum mechanical 
problem for the super-LFT to derive the reflection amplitudes
showing that they are consistent with the quoted results.
In sect.3 we consider the off-critical integrable models 
of the ShG, the BD and the SShG models and compute
the scaling functions using the reflection amplitudes methods
along with the quantization conditions.
We perform the analytic computations in sect. 4 for the various TBA 
equations generalizing the leading order computation upto several higher
orders enough for us to conclude that we can find
nonperturbative equations identical to the quantization conditions.
Numerical confirmations of our results are presented in sect.5
and some relevent discussions are made in sect.6.


\section{Reflection Amplitudes}

In this section, we introduce the reflection amplitudes for the LFT and
super-LFT by quoting references. We interpret the reflection amplitudes
as a reflection of quantum mechanical wave functional of zero-modes.

\subsection{Liouville Field Theory}

The LFT has been studied actively due to its relations to 2D quantum
gravity and string theory and been shown that it enjoys all the
properties as a CFT.
The LFT action defined on a large disk $\Gamma$ of radius $R\to\infty$ is
\[
{\cal A}_L={1\over 4\pi}\int_\Gamma\left[(\partial_a\phi)^2+4\pi\mu e^{2b\phi}\right]
d^2x+{Q\over\pi R}\int_{\partial\Gamma}\phi dl+2Q^2\ln R
\]
where $b$ is the dimensionless Liouville coupling constant and the scale
parameter $\mu$ is usually called the cosmological constant and
the background charge $-Q$ at infinity is
\[
Q=b+1/b.
\]
The LFT is a CFT with central charge
\[
c_L=1+6Q^2
\]
and the dimensions of exponential operators
\[
V_\alpha(x)=e^{2\alpha\phi(x)}
\]
given by
\[
\Delta_\alpha=\alpha(Q-\alpha).
\]
Since $V_{Q-\alpha}$ and $V_{\alpha}$ have the same dimension,
$V_{Q-\alpha}$ is called ``the reflection image'' of $V_\alpha$ and
vice versa. These two operators are dual to each other.

The $n$-point correlation function of the exponential fields
defined as a functional integral
\[
{\cal G}_{\alpha_1,\ldots,\alpha_n}(x_1,\ldots,x_n)=\int V_{\alpha_1}(x_1)
\ldots V_{\alpha_n}(x_n)e^{-A_L[\phi]}D\phi
\]
can be determined as a formal expansion of the cosmological constant
and computed exactly.
The result shows an interesting relation between the correlation
functions ${\cal G}_{\alpha_1,\ldots,\alpha_n}$ and
${\cal G}_{Q-\alpha_1,\ldots,\alpha_n}$.
As an example, consider the three-point function
which can be written as
\[
{\cal G}_{\alpha_1,\alpha_2,\alpha_3}(x_1,x_2,x_3)=
|x_{12}|^{2\ga_{3}}|x_{23}|^{2\ga_{1}}|x_{31}|^{2\ga_{2}}
C(\alpha_1,\alpha_2,\alpha_3)
\]
where $C(\alpha_1,\alpha_2,\alpha_3)$ is the structure constant
and $\ga_1=\De_{\al_1}-\De_{\al_2}-\De_{\al_3}$ and so on.
There are several independent methods to compute the correlation
functions; functional integral \cite{goulli,dornotto}, 
the canonical treatment
\cite{gervais}, and on-mass-shell condition\cite{ZamZam}.

The reflection $\alpha\to Q-\alpha$ of each of the three operators
introduces the Liouville reflection amplitude $S_{L}(P)$
\beq
C(Q-\alpha_1,\alpha_2,\alpha_3)=C(\alpha_1,\alpha_2,\alpha_3)
S_L(i\alpha_1-iQ/2)
\label{strcst}
\eeq
where
\beq
S_{L}(P)=-\left(\pi\mu\gamma(b^2)\right)^{-2iP/b}
{\Gamma(1+2iP/b)\Gamma(1+2iPb)\over\Gamma(1-2iP/b)\Gamma(1-2iPb)}
\label{refamp}
\eeq
with
\[
\gamma(x)=\Gamma(x)/\Gamma(1-x).
\]
By construction, the reflection amplitudes can be also defined from
a two-point function 
\[
\langle V_{\al}(z,{\bar z})V_{\al}(0,0)\rangle=
{S_L(i\al-iQ/2)\over{|z|^{2\De_{\al}}}}\quad
{\rm with}\quad
\langle V_{\al}(z,{\bar z})V_{Q-\al}(0,0)\rangle=
{1\over{|z|^{2\De_{\al}}}}.
\]

Consider LFT on a cylinder of circumference $2\pi$ with the
cartesian coordinates $x_1$, $x_2$ where
$x_2$ along the cylinder is defined as the imaginary time
and $x_1\sim x_1+2\pi$ is the space coordinate.
The Hamiltonian acting in the space of states ${\cal A}$ of LFT
\[
H=-{c_L\over 12}+L_0+\bar L_0
\]
generates translations along the time $x_2$.
The space of states ${\cal A}$ is classified in the highest weight
representations of $Vir\otimes\bar{Vir}$
\[
{\cal A}=\oplus_P{\cal A}_P
\]
where a conformal class ${\cal A}_P$ contains a primary state
$v_P\equiv V_{\al}$ with
\[
\al={Q\over{2}}+iP
\]
and satisfies
\bea
L_n v_P&=&\bar L_n v_P=0\qquad\qquad\qquad {\rm at}\qquad n>0 \\
L_0 v_P&=&\bar L_0 v_P=(Q^2/4+P^2)v_P.
\eea
The primary state $v_P$ corresponding to the exponential operator
$V_{\al}$ becomes the lowest energy state
and its descendants are generated by the action of $L_n$ and $\bar L_n$
with $n<0$ on $v_P$.
Also the terminology of the ``reflection image'' becomes clear since
the operator $V_{Q-\al}$ is represented
by the primary state $v_{-P}$.
Right and left generators $L_n$ and $\bar L_n$ commute and therefore
${\cal A}_P$ has the structure of a direct product of right and
left modules.

In the LFT, one can reformulate the conformal structure in terms of the
``zero-mode'' of the Liouville field $\phi(x)$ defined by
\[
\phi_0=\int_0^{2\pi}\phi(x){dx_1\over 2\pi}.
\]
As $\phi_0\to-\infty$ in the configuration space,
one can neglect the exponential interaction term in the LFT action
so that one can expand $\phi(x)$ as a free massless field
($z=x_1+ix_2$)
\[
\phi(x)=\phi_0-{\cal P}(z-\bar z)+\sum_{n\ne 0}\left({ia_n\over n}e^{inz}+
{i\bar a_n\over n}e^{-in\bar z}\right),
\]
where we defined the momentum conjugate to the zero-mode $\phi_0$ as
\[
{\cal P}=-{i\over 2}{\partial\over\partial\phi_0}
\]
and the oscillators satisfy
\[
\left[a_m,a_n\right]={m\over 2}\delta_{m+n},\quad
\left[\bar a_m,\bar a_n\right]={m\over 2}\delta_{m+n}.
\]
The Virasoro generators are given by
\beaq
L_n&=&\sum_{k\neq 0,n}a_k a_{n-k}+(2{\cal P}+inQ)a_n,\qquad n\neq 0
\nonumber\\
L_0&=&2\sum_{k>0}a_{-k}a_k+Q^2/4+{\cal P}^2,\label{viragen0}
\eeaq
and similarly for ${\bar L}_{n}$'s.
The space of states is now represented as
\beq
{\cal A}_0={\cal L}_2(-\infty<\phi_0<\infty)\otimes{\cal F}
\label{space}
\eeq
where ${\cal L}_2$ is the two-dimensional phase space spanned by
$\phi_0$ and its conjugate momentum ${\cal P}$ and
${\cal F}$ is the Fock space of the oscillators.

Any state $s\in{\cal A}$ can be represented by a wave functional
$\Psi_s[\phi(x_1)]$ in the $\phi_0\to-\infty$ asymptotic limit.
In particular, the wave functional for the primary state $v_P$ corresponds to
\beq
\Psi_{v_P}[\phi(x_1)]=\left(e^{2iP\phi_0}+S(P)e^{-2iP\phi_0}\right)
\left|0\right>\ \ \ \ {\rm as}\ \ \ \phi_0\to-\infty
\label{wavefct1}
\eeq
where $S(P)$ is the reflection coefficient of the asymptotic wave functional.
One can check that the wave functional of asymptotic form Eq.(\ref{wavefct1})
has correct conformal dimension by acting $L_0$ in
Eq.(\ref{viragen0}).
The coefficient $S(P)$ should be the reflection amplitude $S_{L}(P)$
introduced earlier since the wave functional $\Psi_{v_{-P}}$ for the
primary state $v_{-P}$ should be $S_{L}(-P)\Psi_{v_P}$ to be
consistent with Eq.(\ref{strcst}) along with
\[
S_{L}(P)S_{L}(-P)=1.
\]

In this framework, one can check the validity of the reflection amplitude
by taking semiclassical limit $b\to 0$ and using duality.
Since $P$ is of the order of ${\cal O}(b)$,
one can neglect the oscillators and keep only the zero-mode
$\phi_0$ so that the Hamiltonian is approximated as
\[
H_0=-{1\over 12}-{1\over 2}{\partial^2\over\partial\phi_0^2}+2\pi\mu
e^{2b\phi_0}.
\]
The exact wave function of $\phi_0$ for the Hamiltonian is well-known
whose asymptotic form as $\phi_0\to-\infty$ is given by 
Eq.(\ref{wavefct1}) with
\[
S(P)=-\left({\pi\mu\over b^2}\right)^{-2iP/b}{\Gamma\left(1+2iP/b\right)\over
\Gamma\left(1-2iP/b\right)}.
\]
It is straightforward to check this result is consistent with the
non-perturbative reflection amplitude Eq.(\ref{refamp}) perturbatively.

\subsection{super-Liouville Field Theory}

Now we extend above formalism to the $N=1$ super-LFT
whose lagrangian is given by
\[
{\cal L}_{\rm SL}={1\over{8\pi}}(\pd_{a}\phi)^2
- \frac{1}{2\pi}(\bar\psi \pd \bar\psi + \psi \bar\pd \psi)
+i\mu b^2 \psi \bar\psi e^{b\phi} 
+{\pi\mu^2 b^2\over{2}} e^{2b\phi} 
\]
With the background charge $Q$
\[
Q=b+1/b
\]
the central charge of the super-LFT is
\[
c_{SL}={3\over{2}}(1+2Q^2).
\]
The (NS) primary fields of the super-LFT are given by
\[
V_{\alpha} (z,\bar{z},\th,\bar{\th})=\phi_\alpha (z,\bar{z})+\theta
\psi_\alpha (z,\bar{z})+
\bar{\theta}\bar {\psi}_\alpha (z,\bar{z})-\theta\bar{\theta}
\tilde{\phi}_\alpha (z,\bar{z})
\]
with dimensions
\[
\De_{\al}={1\over{2}}\al(Q-\al),
\]
and the (R) fields by
\[
\sigma ^{(\epsilon)}V_{\al}
\]
where $\sigma^{(\epsilon)}$ is the `twist field' with dimension $1/16$
so that the dimension of the (R) fields are
\[
\Delta_{\al}=\frac{1}{16}+{1\over{2}}\al(Q-\al).
\]

The reflection amplitudes of the super-LFT defined from the structure
constants have been derived from the structure constants in
\cite{Rash,Pogho}.\footnote{
We quote the results after some minor corrections in such a way that
they are consistent with both classical results and TBA.}
The reflection amplitudes for the (NS) fields are
\beq
S_{NS}(P)=-\left(\frac{\pi \mu }{2}\gamma \left(\frac{1+b^2}{2}\right)
\right)^{-\frac{2iP}{b}}
\frac{\Gamma (1+iPb)\Gamma\left(1+\frac{iP}{b}\right)}
{\Gamma (1-iPb)\Gamma\left(1-\frac{iP}{b}\right)},
\label{refns}
\eeq
and for the (R) fields
\beq
S_R(P)=\left(\frac{\pi \mu }{2}\gamma \left(\frac{1+b^2}{2}\right)
\right)^{-\frac{2iP}{b}}
\frac{\Gamma \left(\frac{1}{2}+iPb\right)
\Gamma \left(\frac{1}{2}+\frac{iP}{b}\right)}
{\Gamma \left(\frac{1}{2}-iPb\right)\Gamma
\left(\frac{1}{2}-\frac{iP}{b}\right)}.
\label{refr}
\eeq

The super-LFT is a super-CFT which satisfies the
usual super-Virasoro algebra.
The space of states for the super-LFT can be expressed by
\beq
{\cal A}_0={\cal L}_2(-\infty<\phi_0<\infty,\psi_0)\otimes{\cal F}
\label{sspace}
\eeq
where the fermionic zero-mode appears only for the (R) sector and
${\cal F}$ is the Fock space of bosonic and fermionic oscillators.
The zero modes appear in the super-Virasora generator $L_0$ and $S_0$ 
of the (R) sector in such a
way that $L_0$ contains the square of the conjugate momentum ${\cal P}$
like Eq.(\ref{viragen0}) and $S_0$ acts non-trivially only on the
twist field.

The primary state $v_{P}$ 
can be also expressed by a wave functional $\Psi_{v_{P}}[\phi(x_1)]$
whose asymptotic form is given similarly as Eq.(\ref{wavefct1}).
The amplitude $S(P)$ is either $S_{NS}(P)$ or $S_{R}(P)$ depending on
the sector so that the wave functional $\Psi_{v_{-P}}$ is given by
$S(-P)\Psi_{v_P}$.

One can also check the validity of this expression by taking the classical
limit of $b\to 0$.
Since $P$ is small of order of ${\cal O}(b)$,
one can neglect the oscillator part in Eq.(\ref{sspace}) and study only the
dynamics of zero-modes. In the (NS) sector, only bosonic
zero-mode appears so that the Hamiltonian becomes
\[
H_0^{\rm NS} = -\frac18 -\left(\frac{\pd}{\pd\phi_0}\right)^2
+\pi^2\mu^2 b^2 e^{2b\phi_0},
\]
which is essentially the same as that of the LFT, hence the
reflection amplitude becomes
\[
S_{\rm NS}(P)=-\left(\frac{\pi\mu}{2}\right)^{-\frac{2i}{b}P}
         \frac{\Gamma(1+iP/b)}{\Gamma(1-iP/b)}.
\]

On the other hand, in the (R) sector, additional fermionic zero-mode
is introduced in the hamiltonian by \cite{CG}
\[
H_0^{\rm R} = - \left(\frac{\pd}{\pd\phi_0}\right)^2
             + \pi^2\mu^2 b^2 e^{2b\phi_0}
             + 2\pi i \mu b^2 \psi_0 \bar\psi_0 e^{b\phi_0}.
\]
Since the fermionic zero-mode satisfies
\[
\{\psi_0,\bar\psi_0\}=0,\qquad \psi_0^2=\bar\psi_0^2={1\over{2}},
\]
we can represent it by
\[
\psi_0={1\over{\sqrt{2}}}\si_1,\quad
\bar\psi_0={1\over{\sqrt{2}}}\si_2,\quad
\psi_0\bar\psi_0={i\over{2}}\si_3,
\]
and the Hamiltonian becomes
\[
H_0^{\rm R}={\cal P}^2+\pi^2\mu^2 b^2 e^{2b\phi_0}-
\pi \mu b^2 e^{b\phi_0}\si_{3}=\left(
\begin{array}{cc}
H_{+}&0\\
0&H_{-}
\end{array}
\right).
\]
Among two eigen-spinors, the lower energy-state 
can be obtained as
\[
\Psi_{+}(\phi_0)=\left(
\begin{array}{c}
\sqrt{x}\left[K_{1/2-iP/b}(x)+K_{1/2+iP/b}(x)\right]\\
0 \end{array}\right),\quad x=\pi\mu e^{b\phi_0}
\]
where $K_{\nu}(x)$ is the modified Bessel function.
By taking the asymptotic limit $\phi_0\to-\infty$, one can find
the non-vanishing component is given by 
\[
\Psi\sim e^{iP\phi_0}+S_{R}(P)e^{-iP\phi_0}
\]
with
\[
S_{\rm R}(P) =\left(\frac{\pi\mu}{2}\right)^{-\frac{2i}{b}P}
         \frac{\Gamma(\frac12+iP/b)}{\Gamma(\frac12-iP/b)}.
\]
These are consistent with the exact result Eq.(\ref{refr}) 
in the $b\to 0$ limit.


\section{Scaling Functions from Quantization Conditions}

We consider the ShG, the BD and the SShG models as integrable 
perturbations of the LFT and the super-LFT and compute the 
scaling functions by introducing the quantization conditions
for the conjugate momentum in such a way that one can relate
the momentum to the scaling parameter $R$ through the reflection
amplitudes.

\subsection{Perturbations of the LFT}

\leftline{\underline{\bf the ShG model}}

We start by reviewing the analysis of \cite{ZamZam} for the ShG model
or $A_{1}$ affine Toda field theory defined first on a circle of
circumference $R$ with periodic boundary condition.
By rescaling the size to $2\pi$, one can write the action as
\beq
{\cal A}_{\rm ShG}=\int dx_2\int_0^{2\pi}dx_1\left[{1\over 4\pi}
\left(\partial_a\phi\right)^2+\mu\left({R\over 2\pi}\right)^{2+2b^2}
\left(e^{2b\phi}+e^{-2b\phi}\right)\right].
\label{actshg}
\eeq
where $\mu\sim [{\rm mass}]^{2+2b^2}$ is the dimensional coupling constant
with $b$ the coupling constant.

We are interested in the ground state energy $E(R)$ or, more conveniently, the
finite-size effective central charge
\beq
c_{\rm eff}(R)=-{6R\over\pi}E(R)
\label{casimir}
\eeq
in the ultraviolet limit $R\to 0$.
Since we are interested in the ground-state energy, 
only the zero-mode contribution counts.
So the corresponding effective central charge
at $R\to 0$ is determined mainly by $P$
\beq
c_{\rm eff}(R)=1-24P^2+{\cal O}(R)
\label{ceffective}
\eeq
up to power corrections in $R$.

For the ground state energy, one can consider only the zero-mode 
dynamics where the wave functional of $\phi_{0}$ is confined in the 
potential barrier due to the ShG interaction term.
The ShG potential introduces a quantization condition for
the momentum $P$ which depends on the finite size $R$.
As $R\to 0$, in particular, the wave functional is confined in the
potential well where the potential vanishes in the most of the region
and becomes nontrivial at $2b\phi_0\sim\pm\ln\mu(R/2\pi)^{2+2b^2}$ near
the left and right edges.
Near these edges of the potential well, the potential becomes that of the
LFT and the wave functional will be  reflected  with the reflection amplitude
of the LFT introduced earlier.
Therefore, the quantization condition is given by
\[
\left(R/2\pi\right)^{-8iPQ}S_{L}^2(P)=1.
\]
In terms of  the reflection phase $\delta_{L}(P)$ defined by
\[
S_{L}(P)=-e^{i\delta_{L}(P)},
\]
the ground state momentum is qunatized as
\beq
\delta_{L}(P)=\pi+4PQ\ln{R\over{2\pi}}.
\label{quantlft}
\eeq

Thus determined quantized momentum will give
the scaling function $c_{\rm eff}(R)$ in the UV region
by Eq.(\ref{ceffective}).
To see this explcitly, one can  
expand the reflection phase in the odd powers of $P$,
\beq
\delta_{L}(P)=\delta_1(b)P+\delta_3(b)P^3+\delta_5(b)P^5+\ldots
\label{pbc}
\eeq
where the coefficients can be obtained from the reflection amplitude 
Eq. (\ref{refamp}) as follows:
\bea
\delta_1(b)&=&{4\over b}\ln b^2-4Q\ln{\Gamma\left({1\over
2+2b^2}\right)\Gamma\left(1+{b^2\over 2+2b^2}\right)\over 4\sqrt{\pi}}
+\ga_{\rm E}\\
\delta_3(b)&=&{16\over 3}\zeta(3)(b^3+b^{-3})\\
\delta_5(b)&=&{64\over 5}\zeta(5)(b^5+b^{-5})
\eea
with Euler constant $\gamma_{\rm E}$.
Now solving Eq.(\ref{pbc}) iteratively, we get
\beq
c_{\rm eff}(R)=1+{c_1\over{l^2}}+{c_2\over{l^5}}+
{c_3\over{l^7}}+\ldots
\label{lexpand}
\eeq
where
\beaq
l&=&\delta_1(b)-4Q\ln(R/2\pi)\label{lftl}\\
c_1&=&-24\pi^2\label{lftci}\nonumber\\
c_2&=&-48\pi^4\delta_3(b)\nonumber\\
c_3&=&-48\pi^5(2\delta_3(b)^2-\pi\delta_5(b)).\nonumber
\eeaq

The Gamma functions appear in $\delta_1$ due to the relation
between the mass of the physical particle 
and the coupling constant $\mu$ in the action \cite{zammass}
\[
-{\pi\mu\over\gamma(-b^2)}=\left[{m\over 4\sqrt{\pi}}
\Gamma\left({1\over 2+2b^2}\right)\Gamma
\left(1+{b^2\over 2+2b^2}\right)\right]^{2+2b^2}.
\]

\vskip 1cm
\leftline{\underline{\bf the BD model}}

The BD model is an integrable field theory associated
with $A_{2}^{(2)}$ affine Toda theory and can be regarded as an integrable
perturbation of the LFT \cite{FLZZ}.
The action is given on a circle of circumference $2\pi$ with periodic
boundary condition;
\[
{\cal A}_{BD}=\int dx_2\int^{2\pi}_{0} dx_1
\left[{1\over{4\pi}}(\partial_{a}\phi)^2 + \mu
\left({R\over{2\pi}}\right)^{2+2b^2}e^{2b\phi} +
\mu'\left({R\over{2\pi}}\right)^{2+b^2/2} e^{-b\phi}\right]
\]
where $\mu\sim[{\rm mass}]^{2+2b^2}$ and $\mu'\sim[{\rm mass}]^{2+b^2/2}$
are related to the mass of on-shell particle by
\[
m={2\sqrt{3}\Gamma\left({1\over 3}\right)\over{
\Gamma\left(1+{b^2\over 6+3 b^2}\right)\Gamma\left({2\over 6+3 b^2}\right)}}
\left[-{\mu\pi\Gamma\left(1+b^2\right)\over{
\Gamma\left(-b^2\right)}}\right]^{{1\over  6+3 b^2}}
\left[- {2{\mu'}\pi\Gamma\left(1+{ b^2\over 4}\right)\over{
\Gamma\left(-{ b^2\over 4}\right)}}\right]^{{2\over  6+3 b^2}}.
\]

This model possesses asymmetrical exponential
potential terms compared with the ShG model.
In the UV limit, the exponential potential becomes negligibly small
except in the region where $\phi_{0}$ goes to $\pm\infty$.
This means that the BD model is again effectively described by
the LFT.
The scaling function of the central charge is
given by the same Eqs.(\ref{casimir}) and (\ref{ceffective})
in the ShG model.
It is the quantization condition that makes the
difference from the ShG model,
due to the asymmetry of the potential well in the left and right edges.
The conjugate momentum $P$ is now quantized by the condition
\[
\left({R\over{2\pi}}\right)^{-4iP(Q+Q')}S_{L}(P)S_{L}'(P)=1,
\]
where $S_{L}'(P)$ is obtained by substituting $b\to b/2$ for $S_{L}(P)$
given in Eq.(\ref{refamp}) and
\[
Q=b+1/b,\quad Q'=b/2+2/b.
\]
Using the phase shifts defined as
\[
S_{L}(P)=-e^{i\de_{L}(P)},\qquad S_{L}'(P)=-e^{i\de_{L}'(P)},
\]
the quantization condition becomes
\beq
\bar\de(P)=\pi+4\bar{Q}P\ln{R\over{2\pi}}
\label{quantize}
\eeq
where
\[
\bar\de(P) = \frac12(\de_L(P)+\de'_L(P)), \quad \bar{Q} = \frac12(Q+Q').
\]

The relation between $P$ and $R$ in Eq.(\ref{quantize})
gives the scaling function $c_{\rm eff}$ as a continuous function of $R$,
Eq.(\ref{lexpand}), with $Q$ replaced by $\bar Q$
and $\delta$'s with $\bar\de$'s defined by
power series expansion of the phase shift in $P$ 
\beaq
\label{bd_delta}
\bar\de(P)&=&\bar\de_{1}P+\bar\de_{3}P^3+\bar\de_{5}P^5+\ldots\nonumber\\
\bar\de_{1}&=&{6\over{b}}\ln{b^2\over{2}}-2(Q+Q')\left[
\ln{m\Gamma\left(1+{b^2\over 6+3 b^2}\right)
\Gamma\left({2\over 6+3 b^2}\right)\over{2\sqrt{3}
\Gamma\left({1\over 3}\right)}}+\gamma_{\rm E}\right]\nonumber\\
\bar\de_{3}&=&3\zeta(3)\left(b^3+{8\over{b^3}}\right)\nonumber\\
\bar\de_{5}&=&-{33\over{5}}\zeta(5)\left(b^5+{32\over{b^5}}\right).
\eeaq

\subsection{the SShG model}

Now we consider an integrable model obtained as a perturbation of
the super-LFT, the SShG model.
By rescaling the size to $2\pi$, 
one can express the action of the SShG model by
\bea
{\cal A}_{\rm SShG}&=&\int dx_2\int_{0}^{2\pi}dx_1\left[
{1\over{8\pi}}(\pd_{a}\phi)^2
-\frac{1}{2\pi}(\bar\psi \pd \bar\psi + \psi \bar\pd \psi)
\right.\nonumber \\
&+&\left.2i\mu b^2\left({R\over{2\pi}}\right)^{1+b^2}
\psi \bar\psi\cosh(b\phi)+
{\pi\mu^2 b^2}\left({R\over{2\pi}}\right)^{2+2b^2}
\left[\cosh(2b\phi)-1\right] \right].
\eea
In the UV limit, the exponential potential becomes negligible
except in the region where $\phi_{0}$ goes to $\pm\infty$.
This means that the SShG model is effectively described by
the super-LFT as $R\to 0$.  From the ground state energy
for the primary state labelled by
$P$, the effective central charge can be obtained by
\bea
c_{\rm eff}(R)&=&{3\over{2}}-12P^2+{\cal O}(R)\qquad{\rm (NS)}\nonumber\\
&=&-12P^2+{\cal O}(R)\qquad{\rm (R)}.
\eea

For the (NS) sector, $P$ corresponding to the ground state is 
determined again by the
quantization condition coming from the super-LFT reflection amplitudes:
\beq \label{squantize}
\de_{\rm NS}(P)=\pi+2QP\ln{R\over{2\pi}},
\eeq
where $\de_{\rm NS}(P)$ is the phase factor of (NS) reflection amplitudes. 
This quantization condition can be solved iteratively
by expanding $\de_{\rm NS}(P)$ in powers of $P$, 
\beaq \label{ns_phase}
\de_{\rm NS}(P) &=& \de_1^{\rm NS} P + \de_3^{\rm NS} P^3
                  + \de_5^{\rm NS} P^5 + \cdots \nonumber \\
\de_1^{\rm NS} &=& -2\left\{
\frac1b \ln\left[{\frac{\pi\mu}2} \ga\left(\frac{1+b^2}2\right)\right]
    + \ga_{\rm E}Q   
\right\} \nonumber\\
\de_3^{\rm NS}&=&\frac23\zeta(3)\left(b^3+\frac1{b^3}\right)\nonumber\\
\de_5^{\rm NS} &=& -\frac25\zeta(5) \left( b^5 + \frac1{b^5} \right).
\eeaq
To decide the phases completely, one needs a relation between
$\mu$ and $m$ for the SShG model. 
This is given in \cite{ssgmass} by
\beq
{\pi\over{2}}\mu b^2\ga\left({1+b^2\over{2}}\right)=
\left[{m\over{8}}{{\pi b^2\over{1+b^2}}\over{\sin
{\pi b^2\over{1+b^2}}}}\right]^{1+b^2}.
\label{ssgmmu}
\eeq
In terms of these coefficients, one can find the scaling function in
a similar way as Eq.(\ref{lexpand}). 

We will consider the (R) sector in the next section 
since there is a fundamental difference from the (NS).


\section{Perturbative TBA analysis in the UV region}

We compute the scaling functions analytically in the deep UV region
by extending the methods developed in \cite{roaming,martins,fring}
to higher orders.
We find a hidden connection between the TBA and the quantization 
conditions arising from the reflection amplitudes.

\subsection{the ShG and the BD models}

The TBA equations for the ShG and the BD models are given by
($r\equiv mR$)
\beq
r\cosh\theta=\ep(\theta)+\int\varphi(\theta-\theta')\ln\left(1+
e^{-\ep(\theta')}\right){d\theta'\over 2\pi},
\label{shgtba}
\eeq
where the scaling function is expressed with the `pseudo-energy' 
$\ep(\theta)$ by
\beq
c_{\rm eff}(R)={3r\over\pi^2}\int\cosh\theta\ln\left(1+e^{-\ep(
\theta)}\right)d\theta \,.
\label{shgcent}
\eeq
The model dependence comes from the kernel. 
The kernel of the ShG model is given by 
\[
\varphi(\th)=-\Phi_{B}(\th),
\]
where
\[
\Phi_{x}(\th)\equiv{4\sin\pi x\cosh\th\over{\cos 2\pi x-\cosh 2\th}},
\quad {\rm with}\quad B={b^2\over{1+b^2}}
\]
and for the BD model 
\[
\varphi(\th)=\Phi_{2/3}(\th)+\Phi_{-B/3}(\th) +\Phi_{(B-2)/3}(\th),
\qquad {\rm with}\quad B={b^2\over{1+b^2/2}}.
\]

Following \cite{roaming}, we express the effective central charge as
\[
c_{\rm eff}(r)={6\over{\pi^2}}\int_{0}^{\infty}r\cosh\th L(\th,r)d\th.
\]
by defining
\[
L(\th,r)=\ln(1+e^{-\ep(\th,r)}).
\]
Fourier transform of the kernel $\varphi(\th)$
\[
{\tilde\varphi}(k)=\int_{-\infty}^{\infty}d\th\varphi(\th)e^{ik\th}
=2\pi\left\{ 1 + \sum_{n=0}^{\infty}(-i)^{n}\phit_n k^{n} \right\}
\]
rewrites Eq.(\ref{shgtba}) as an infinite order differential equation
\[
r\cosh\th+\ln\left(1-e^{-L(\th,r)}\right)=
\sum_{n=0}^{\infty}\phit_n L^{(n)}(\th,r),
\]
where
\[
L^{(n)}(\th,r)=(d/d\th)^{n}L(\th,r).
\]

We are going to concentrate on positive rapidity ($\th>0$)
since the TBA is even in $\th$.
One can extend the solution to negative $\th$ using this symmetry.
In the UV limit ($r\to 0$), since 
$r\cosh\th $ can be approximated as $re^{\th}/2$,
it is convenient to rescale $\th$ and define a new function
\[
{\hat L}(\th)\equiv L(\th-\ln(r/2),r)
\]
so that the TBA equation becomes
\beq
e^{\th}+\ln\left(1-e^{-{\hat L}(\th)}\right)=
\sum_{n=0}^{\infty}\phit_n{\hat L}^{(n)}(\th).
\label{newtba}
\eeq
The central charge
\[
c_{\rm eff}(r)={6\over{\pi^2}}\int_{\ln(r/2)}^{\infty}
e^{\th}{\hat L}(\th)d\th
\]
becomes, after integrating by parts using $\frac{d}{d\th}e^\th = e^\th$,
\bea
c_{\rm eff}(r)&=&{3\over{\pi^2}}\left(\sum_{n=1}^{\infty}\phit_{2n}
\sum_{k=1}^{2n-1}(-1)^{k+1}{\hat L}^{(k)}(r'){\hat L}^{(2n-k)}(r')+
\phit_{0}{\hat L}(r'){\hat L}(r')\right)\\
&-&{6\over{\pi^2}}\left(-{\cal L}(1-e^{-{\hat L}(r')})+\half
{\hat L}(r')\ln(1-e^{-{\hat L}(r')})+e^{r'}{\hat L}(r')\right),
\eea
where $r'\equiv \ln{(r/2)}$ and ${\cal L}$ is the Rogers dilogarithm
function. 

To solve the rescaled TBA, Eq.~(\ref{newtba}),
we neglect the driving term $e^\th$ and regard $e^{- \hat L(\th)}$
as the same order of $\hat L^{(n)}$.
(We are solving $\hat L (\th)$ around the plateau region
since $c_{\rm eff}$
is given in terms of $\hat L^{(k)} (r')$'s).
The leading order of the TBA becomes the Liouville equation,
\[
\phit_{2}{\hat L}_0^{(2)}(\th)+e^{-{\hat L}_0(\th)}=0\,,
\]
whose solution is given by
\beq
{\hat L}_0(\th)=\ln\left({\sin^2(\al(\th-\be))\over{2\al^2\phit_{2}}}\right)
\label{tbasoli}
\eeq
and the  lowest order of the effective central charge is given by
\[
c_{\rm eff}(r) =1+{3\over{\pi^2}}\left(\phit_{2}{\hat L}^{(1)}(r')^2
-2e^{-{\hat L}(r')}\right)+\cdots=1-{12\over{\pi^2}}\phit_{2}\al^2
+ \cdots\,.
\]

$\al$ and $\be$ are integration constants,
which are to be fixed by additional input.
We first note that
due to the symmetry property
$L(\th,r)=L(-\th,r)$ or
\beq
\label{sym_L}
{\hat L} (\th) = {\hat L} (2\ln(r/2)-\th)\,,
\eeq
an important relation appears  between $\al$ and $\be$
\beq
2\al(\be-\ln{(r/2)}) = n \pi,
\label{relationi}
\eeq
where $n$ is an arbitrary odd integer and is fixed as $n=1$.
Using this relation, we can reexpress Eq.(\ref{tbasoli}) as
\beq
{\hat L}_0(\th)=\ln\left[{\cos^2(\al(\th-\ln(r/2)))
\over{2\al^2\phit_{2}}}\right].
\label{tbasolii}
\eeq

As $r \to 0$, $\be$ remains finite while $\al \to 0$.
In addition, $\al$ should be small, $\al^2 \phit_{2} << 1$
to make $\hat L_0 >0$.
Still, $\al$ and $\be$ are  not completely fixed at this stage.
We need the correct solution which
vanishes as $\th \to \infty$ (tail part of $\hat L$)
to fix the integration constant.
The lowest solution $\hat L_0 (\th)$ does
not satisfy the correct condition, since
we restrict the solution on the plateau:
$\th$ may be restricted in the region
\[
0 < \al ( \th - \ln{(r/2)}) < {\pi\over{2}}-\sqrt{2 \alpha^2\phit_{2}}
\]
such that $\hat L_0 (\th)$ is positive and decreases as $\th$ increases.
The restriction of the rapidity domain does not
introduce much errors in the $c_{\rm eff}$:
the error is the order of ${\cal O}(r)$,
\[
c_{\rm eff} \sim {6r \over \pi^2}
\int_{{\pi \over 2 \alpha} - \sqrt{2 \phit_{2}} }^{\infty}
e^\theta L(\theta) d\theta
\sim {\cal O}(r).
\]
Since it is hard to get the complete analytic
solution on the whole rapidity,
we may resort to other physical solution to fix $\al$ and $\be$.
Before doing this, we solve TBA on the plateau  perturbatively.

The rescaled TBA equation (\ref{newtba}) can be solved by 
expanding in a series 
\[
{\hat L}= \sum_{n=0}^\infty {\hat L}_n.
\]
Since ${\hat L}_0^{(n)}(\th)$ and $e^{-n{\hat L}_0(\th)}$
are of the same order,
the expansion can be regarded as the expansion in $\al$.
We give the explicit solution of the TBA up to the order of $\al^8$.
The differential equations for $\hat L_2$, $\hat L_4$, and $\hat L_6$
are given as
\bea
&&\phit_{2} \hat L_2^{(2)} + \phit_{4} \hat L_0^{(4)}
=e^{-\hat L_0} \hat  L_2 - {1 \over 2} e^{- 2 \hat L_0}\\
&&\phit_{2} \hat L_4^{(2)} + \phit_{4} \hat L_2^{(4)}
+ \phit_{6} \hat L_0^{(6)}
=e^{-\hat L_0}\left(\hat  L_4 - {1 \over 2}\hat L_2^2\right)
+ \hat L_2  e^{- 2 \hat L_0}
-{ 1 \over 3}   e^{- 3 \hat L_0}\\
&&\phit_{2} \hat L_6^{(2)} + \phit_{4} \hat L_4^{(4)}
+ \phit_{6} \hat L_2^{(6)} + \phit_{8} \hat L_0^{(6)}
=e^{-\hat L_0}\left( \hat  L_6 - \hat L_2 \hat L_4 
+ \frac16 \hat L_2^3\right) \\
&&\hspace{67mm} + \frac12 e^{-2\hat L_0}\left( 
\hat L_4 - {1 \over 2} \hat L_2^2\right)
+\frac13 \hat L_2  e^{- 3 \hat L_0}
-{ 1 \over 4}   e^{- 4 \hat L_0}\,.
\eea

These equations are iterative in the sense that one can find
the solution for $\hat L_{2n}$ by inserting the solutions of
previous differential equations, $\hat L_{2k}$, $k=1,\ldots n-1$.
Since these equations are inhomogenious, the solutions are 
linear combinations of special solutions and solutions to the 
homogeneous equations
\[
\phit_{2}\hat L''_{2n}=e^{-\hat L_0}\hat L_{2n}.
\]
The solutions to the homogeneous equation are given by 
\[
\hat L_{2n}=c_0(1+x\tan x)+c_1\tan x,\quad x=\al(\th-\ln(r/2)).
\]
The second term $c_1$ should vanish due to the symmetric property 
of $\hat L$ and the first term can be absorbed into $\hat L_0$ 
by redefining constant $\al$.
Therefore, it is enough to consider only special solutions.

The special solutions are surprisingly simple and given
in terms of derviatives of $\hat L_0$,
\bea
&&\hat L_2 = a_{(2,2)} \hat L_0^{(2)} + \al^{2} a_{(2,0)}\\
&&\hat L_4 = a_{(4,4)} \hat L_0^{(4)} + \al^2 a_{(4,2)} \hat L_0^{(2)}
 + \al^4 a_{(4,0)} \\
&&\hat L_6 = a_{(6,6)} \hat L_0^{(6)} + \al^2 a_{(6,4)} \hat L_0^{(4)}
 + \al^4 a_{(6,2)} \hat L_0^{(2)} + \al^6 a_{(6,0)}
\eea
where $a_{(ij)}$'s are constants fixed  by $\phit_{2n}$'s.
\bea
&&a_{(2,2)} = {\phit_{2} \over 4} - {3 \phit_{4} \over 2 \phit_{2}} \\
&&a_{(2,0)} = {\phit_{2} } - {2 \phit_{4} \over  \phit_{2}} \\
&&a_{(4,4)} = {1 \over{2592\phit_{2}^2}}\left[{59 \phit_{2}^4} 
- 900 \phit_{4}\phit_{2}^2 + 4428\phit_{4}^2 -2880 
\phit_{6}\phit_{2}\right]\\
&&a_{(4,2)} = {1 \over{81\phit_{2}^2}}\left[\phit_{2}^4
- 18 \phit_{4}\phit_{2}^2 + 108\phit_{4}^2 
-90 \phit_{6} \phit_{2}\right]\\
&&a_{(4,0)} = {5 \over 27} {\phit_{2}^2} - {4 \over 3}  \phit_{4}
+ 4\left({\phit_{4} \over \phit_{2} }\right)^2
- {8 \over 3}  { \phit_{6} \over \phit_{2}}\\
&&a_{(6,6)} = \frac1{259200\phit_{2}^3} \left[
     281\phit_{2}^6 - 7794\phit_{2}^4\phit_{4} - 468504\phit_{4}^3
    - 69840\phit_{2}^3\phit_{6} \right.\\
&&\hspace{40mm} \left.+ 712800\phit_{2}\phit_{4}\phit_{6}
    + 756\phit_{2}^2 (131\phit_{4}^2 - 360\phit_{8})\right]\\
&&a_{(6,4)} = \frac1{3240\phit_{2}^3}
\left[ 7\phit_{2}^6 - 198 \phit_{2}^4\phit_{4}-9288\phit_{4}^3\right. \\
&&\hspace{40mm} \left.- 1080\phit_{2}^3\phit_{6}
   + 10800 \phit_{2} \phit_{4} \phit_{6}
   + 72\phit_{2}^2 ( 31\phit_{4}^2 - 35\phit_{8})\right]\\
&&a_{(6,2)} = \frac{2}{225\phit_{2}^3}\left[
   \phit_{2}^6 - 24\phit_{2}^4\phit_{4} - 984\phit_{4}^3
  - 190\phit_{2}^3\phit_{6}\right. \\
&&\hspace{40mm} \left. + 1500\phit_{2}\phit_{4}\phit_{6}
  + 16\phit_{2}^2 ( 16\phit_{4}^2 - 35\phit_{8}) \right] \\
&&a_{(6,0)} = \frac{4}{45\phit_{2}^3}\left[
   \phit_{2}^6 - 14\phit_{2}^4\phit_{4} - 264\phit_{4}^3
  - 60\phit_{2}^3\phit_{6}\right. \\
&&\hspace{40mm} \left. + 360\phit_{2}\phit_{4}\phit_{6}
  + 24\phit_{2}^2 ( 4 \phit_{4}^2 - 5 \phit_{8} )\right]\,.
\eea

This solution shows some  remarkable properties in connection with
reflection amplitude.
First, the relation between $\al$ and $\be$
in Eq.~(\ref{relationi})
is not changed after this higher order correction,
since it  is due to the reflection-symmetry property of the solution
Eq.~(\ref{sym_L})
and  the higher order solutions are given in terms of
derivatives of $\hat L_0$.
This symmetry relation
has the exact same analogue in the reflection amplitude, namely, the
quantization condition Eq.(\ref{quantlft}).
Indeed, Eq.~(\ref{relationi}) turns out to be the quantization condition
once the integration constants $\al$ and $\be$ are identified
with $P$ and $\de$ by
\bea
2PQ &=& \al\,,\nonumber\\
\de(P) &=& 2\al\be + 2 \al \ln\pi = \pi + 2\al \ln{(r/2\pi)}\,.
\eea
By setting the mass $m=1$, we will identify $r$ with $R$.

Now one immediately notices  that $\al$ has the role of $P$ and $\be$
the phase of the reflection amplitude
and $\al$ and $\be$ have a new nonlinear relation
in addition to the symmetry condition
Eq.~(\ref{relationi}),
\beq
\be
= - \ln \pi + {\de(P)\over 2\al}
= -\ln \pi
+ { 1 \over 2\al}\left[\left({\alpha \over 2 Q}\right)\de_1 
+\left({\al \over 2Q}\right)^3 \de_3
+ \cdots\right]. 
\label{be_al}
\eeq

Second, the higher order solution gives null contribution
to $c_{\rm eff} (r)$ up to this order,
\beq
c_{\rm eff}(r)
=1-{12\over{\pi^2}}\phit_{2}\al^2
+ {\cal O}(\al^{10})
\label{trunceff}
\eeq
independent of the  details of the  kernel $\varphi$.
We need only $ \phit_{0} =\phit_{2n+1}=0$
and explicit values of $\hat L^{(2n)}$'s at $\th = r'$:
$\hat L_0^{(2)}(r') = -2 \al^2$,
$\hat L_0^{(4)}(r') = -4 \al^4$,
$\hat L_0^{(6)}(r') = -32 \al^6$.
So, Eq.(\ref{trunceff}) becomes exactly  Eq.(\ref{lexpand})
since $\phit_{2} = \pi^2/2Q^2$ in the ShG model.
The same holds for the BD model by replacing
$Q$ with $\bar{Q}$ since $\phit_{2} = \pi^2/2\bar Q^2$.

The fact that higher order corrections vanish upto these orders
makes it very plausible to conjecture that all the corrections indeed
vanish so that $c_{\rm eff}$ has only the lowest order contribution:
\[
c_{\rm eff} (r) = 1 -{12\over{\pi^2}}\phit_{2}  \al^2  
\]
and consistent with the quantization condition arising in the 
reflection amplitudes.
The scaling comes in through
the relations between $\al$ and $\be$ only,
Eqs.~(\ref{relationi}) and (\ref{be_al}).
This is consistent with the reflection amplitude consideration.

\subsection{the SShG model}

The particle specturm of the SShG model is a doublet of mass
degenerate which form a supermultiplet.
Their $S$-matrix has been obtained from the Yang-Baxter equation
\cite{ShaWit} which includes one unknown parameter.
This parameter has been related to the coupling constant
of the SShG model by interpreting the supermultiplet as bound
states (`breathers') of the solitons of the supersymmetric sine-Gordon
model \cite{Ahni}.

Since the conventional TBA analysis gives the effective central
charge in which only the lowest conformal dimension of the theory enter,
the TBA equation will give only the (NS) result.
Explicit derivation of the (NS) TBA based on the non-diagonal
$S$-matrix has been done in \cite{Ahnii} by diagonalizing
the transfer matrix using  the inversion relation.
The scaling function can be expressed by
\beq \label{sshg_c_eff}
c_{\rm eff}(r)={3r\over\pi^2}\int\cosh\th\ln
\left(1+e^{-\ep_{1}(\th)}\right)d\th
\eeq
where the pseudo-energies are the solution of the TBA equation,
\beaq \label{sshg_tba}
\ep_{1}(\th)&=&r\cosh\th-\int{d\th'\over 2\pi}
\varphi(\th-\th')\ln[1+e^{-\ep_{2}(\th')}], \nonumber \\
\ep_{2}(\th)&=&-\int{d\th'\over 2\pi}\varphi(\th-\th')
\ln[1+e^{-\ep_{1}(\th')}]
\eeaq
where $\ga=b^2/(1+b^2)$ and the kernel is that of the ShG model.

Analytic computations in the deep UV region can be done similarly.
A little complication arises due to the coupled TBA
Eq.~(\ref{sshg_tba}).
Defining
\[
\hat L\equiv\ln\left[1+e^{-\ep_1 (\th- \ln{(r/2)},r)}\right]
\quad{\rm and}\quad
\hat M\equiv\ln\left[1+e^{-\ep_2 (\th- \ln{(r/2)},r)}\right]
\]
we have
\bea
e^{\th}&=& -\ln\left[1-e^{-{\hat L}(\th)}\right]
+\hat M - \hat L +
\sum_{n=1}^{\infty}\phit_{2n}{\hat M}^{(2n)}(\th)\\
0&=&  - \ln\left[1-e^{-{\hat M}(\th)}\right]
+\hat L - \hat M +
\sum_{n=1}^{\infty}\phit_{2n}{\hat L}^{(2n)}(\th)\,.
\label{new_stba}
\eea
At plateau ($\th \to - \infty$)
\[
\hat L(r') = \hat M(r')\,,
\]
and at the edge ($\th \to \infty$)
\[
\hat L(\infty) =0\,,\quad\quad
\hat M(\infty) = \ln 2\,.
\]

The effective central charge is given as
\[
c_{NS} (r) = c_0 (r)  + c_k (r)
\]
where
\beaq
c_0 (r)& =& {6 \over \pi^2} \int_{r'}^\infty
\left[\ln(1 - e^{-\hat L})
- \sum \phit_{2n} \hat L^{(2n)}\right]\hat L^{(1)} d \th\nonumber \\
c_k (r)& =& {6 \over \pi^2} \int_{r'}^\infty
\left[\hat L - \hat M - \sum \phit_{2n}
( \hat L^{(2n)} - \hat M^{(2n)})\right]\hat L^{(1)} d \th
\label{ck}
\eeaq
$c_0 (r) $ turns out to be the same as the central charge
of the ShG model.
This is because $\hat L $ vanishes at the edge and
$\hat L(r')$ is described by the same TBA of the ShG model 
Eq.~(\ref{newtba}).

$c_k(r)$ has the dominant contribution from the kink side of
$\hat L(\th)$ (non-vanishing part of $\hat L^{(1)}$).
To express $c_k$ in terms of $\hat L (r')$
we need two  identities.
One is given by integrating  by part and
substituting $\hat M$ with $\hat L$ using Eq.~(\ref{new_stba})
\bea
\int_{r'}^\infty \sum \phit_{2n} \hat M^{(2n)} \hat L^{(1)} d\th
&=& \int_{r'}^\infty
\ln (1 - e^{-\hat M}) \hat M^{(1)} d\th\\
&+&\sum \phit_{2n}\left[\sum_k^{2n-1} (-1)^{(k+1)}
\hat M^{(2n-k)} (r') \hat L^{(k)} (r')\right].
\eea
The other is given in terms of Rogers dilogarithmic function
\[
\int_{r'}^\infty
\ln (1 - e^{-\hat M}) \hat M^{(1)} d\th
- {1 \over 2} \hat M(\infty)^2
={\cal L}\left({1 \over 2}\right)
- \sum_{n=1}^\infty {e^{-n \hat M(r') } \over n^2}\,.
\]
Plugging these two identities into Eq.~(\ref{ck}) we have
\[
c_k(r) =  {1 \over 2} - {3 \over \pi^2}
\left[ \sum_{n,k} \phit_{2n}
(-1)^k \hat L^{(2n-k)}(r') \hat L^{(k)}(r')
+ \sum_n {2 \over n^2} e^{-n \hat L(r')}\right] \,.
\]

Combining $c_k$ with $c_0$ we have the scaling function of the SShG model,
\[
c(r) = {3 \over 2} - {24 \phit_{2} \al_{NS}^2 \over \pi^2} 
+ {\cal O}(\al_{NS}^{10}, r)
\]
where we used the perturbative series results in the ShG model.
$\al_{NS} $ is the integration constant corresponding to $\al$
and $\be_{NS}$ to $\be$.
Now recalling the reflection symmetry Eq.~(\ref{sym_L})
and quantization condition of
the reflection amplitude Eq.~(\ref{squantize}),
\bea
2\al_{NS}  (\be_{NS}  - \ln {(r/2)} ) &=& \pi\\
\de_{NS} (p) -2 QP \ln {r \over 2\pi} &=&\pi
\eea
we have the relation between the integration constants
\bea
\al_{NS} &=& QP\\
2 \al_{NS} \be_{NS} &=& \de_{NS}(P) + 2 QP \ln \pi\,.
\eea
With the help of  $\phit_{2} = \pi^2/(2 Q^2)$ for the SShG model,
the effective central charge is given by
\[
c_{NS} = {3 \over 2} - 12P^2 + {\cal O}(\al_{NS}^{10}).
\]

It is not clear how to derive the (R) sector TBA directly.
Instead we can conjecture the TBA equation by considering slightly
different version of the SShG model.\footnote{
We thank Al. Zamolodchikov for suggesting this idea.}
By replacing the $\cosh$ super-potential with $\sinh$ one, we can still
maintain the supersymmetry and integrability and can still consider 
the model as an integrable perturbation of the super-LFT.
The particle spectrum, however, completely changes.
Instead of mass degenerate of a boson and a fermion, we have only
left- and right-moving massless fermionic modes where the boson becomes 
unstable and decays into two fermions. 
The $S$-matrix between the two modes is the same as the ShG model.
It is straightforward to write down the TBA equations since the
scattering is diagonal.
Now consider the (R) sector by imposing anti-periodic boundary condition. 
Following \cite{Fendley} for the Ising model, 
we can find the TBA of the (R) sector as
\bea
\ep_1(\th)&=&r\cosh\th-\varphi*\ln\left(1-e^{-\ep_{2}}\right)(\th)\\
\ep_2(\th)&=&-\varphi*\ln\left(1-e^{-\ep_{1}}\right)(\th)\,.
\eea
The effective central charge becomes
\[
c_{\rm eff}(r)={3r\over{\pi^2}}\int\cosh
\th\ln\left(1-e^{-\ep_{1}(\th)}\right).
\]
At the UV fixed point, one can express it in terms of Rogers 
dilogarithmic functions as
\[
c_{\rm eff}={6\over{\pi^2}}\sum_{a=1,2} \int_{\ep_a(0)}^{\ep_a(\infty)}
d\ep_a\left[ \ln\left(1-e^{-\ep_a}\right)-{\ep_ae^{-\ep_a}\over{
1-e^{-\ep_a}}}\right]
={6\over{\pi^2}}\sum_{a=1,2}\left[ {\cal L}(y_a)-{\cal L}(x_a)\right]
\]
where variables
\[
x_a=e^{-\ep_a(0)},\qquad y_a=e^{-\ep_a(\infty)}
\]
are solutions of simple algebraic equations 
\bea
x_1&=&1-x_2,\qquad x_2=1-x_1\\
y_1&=&0,\qquad y_2=1.
\eea
One can easily check 
\[
c_{\rm eff}
= {6\over{\pi^2}}\left[-{\cal L}(x_1)+{\cal L}(y_2)-{\cal L}(x_2)
\right]=0
\]
using a well-known identity
\[
{\cal L}(x)+{\cal L}(1-x)={\cal L}(1).
\]

In the UV region, there are only power corrections ${\cal O}(R)$. 
To see this, we rewrite the TBA in terms of $H=\ln{(1 - e^{-\ep})}$ 
at the plateau ($\theta \sim r'$) by
\bea
\ln{ (1 - e^{H_1})} &=&  H_2 + \sum_{n=1}^\infty \phit_{2n} H_2^{(2n)}\\
\ln{ (1 - e^{H_2})} &=&  H_1 + \sum_{n=1}^\infty \phit_{2n} H_1^{(2n)}.
\eea
For simplicity let us consider above equations with the assumption
that $H_1 =H_2=H$.
By solving the leading order equation $\ln(1 - e^{H})=  H$ ($e^H= 1/2$),
we can find higher order corrections around this.
The next order correction satisfies the Hook's equation,
$\phit_2 \De H^{(2)} = - 2 \De H$, whose general solution is given by
$\De H= A \sin\omega(\theta -\be)$ with $\omega^2 = 2/ \phit_2$.
Now this solution should satisfy the symmetry Eq.~(\ref{sym_L}),
\[
2\omega(\be-\ln{(r/2)}) = {\rm (integer)}\times \pi, 
\]
It is impossible to satisfy this condition with finte $\om$ 
as $r\to 0$. 
Therefore, the correction $\De H$ should vanish and
$c_{\rm eff} = 0 + {\cal O}(R)$. 
The same conclusion goes with the general case $H_1 \neq H_2$. 

This result is equivalent to fixing the integer $n$ appearing in 
the quantization condition to zero.
The physical meaning becomes clear if one considers the
$P\to 0$ limit where $S_{R}(P)\to 1$ comparing with $S_{NS}(P)\to -1$. 
While for the (NS) sector $\Psi_{P}\sim 2iP\phi_0$ so that the quantum
number $n$ should be $1$ as in Eq.(\ref{squantize}),
the wave functional for the (R) sector becomes constant 
corresponding to $n=0$.
Therefore, the quantization condition becomes
\[
\de_{\rm R}(P)=2QP\ln{R\over{2\pi}}.
\]
Obvious solution is $P=0$ so that
\[
c_{\rm eff}(R)=0+{\cal O}(R).
\]
In the $b\rightarrow 0$ limit, one can verify this from the (R) sector 
zero-mode dynamics of the SShG model which is governed by the Hamiltonain
\[
H_0^{\rm R} = - \left(\frac{\pd}{\pd\phi_0}\right)^2
             + 4 \pi^2\mu^2 b^2 \sinh^2{b\phi_0}
	     + 4\pi i \mu b^2 \psi_0 \bar\psi_0 \cosh{b\phi_0}.
\]
This is a typical supersymmetric quantum mechanics problem and in general 
there exists a zero-energy ground-state \cite{witten} if the supersymmetry is
not broken. Explicitly, the wavefunction of the state is found to be
\[
\Psi_0(\phi_0) = \left(
\begin{array}{c}
e^{-2\pi\mu \cosh{b\phi_0}}\\
0
\end{array}
\right).
\]
This state is normalizable and its energy is exactly zero. Thus at least
in $b\rightarrow 0$ limit, $c_{\rm eff}$ is exactly zero regardless of $r$
without any power correction.


\section{Numerical Analysis}

In this section, we solve TBA equations numerically and confirm the results
in the previous sections.

First we obtain the scaling function $c_{\rm eff}(r)$ of the BD model 
by solving numerically the TBA equation. Then,
this result can be used to produce the reflection amplitude of LFT
through the quantization condition which relates $R(=r)$ and $P$ 
in the following way. 
In $r \rightarrow 0$ limit, one can neglect the ${\cal O}(r)$ power 
correction in Eq.~(\ref{ceffective}) and define $P(r)$ in the TBA framework as
\beq \label{p_c_eff}
P = \sqrt{\frac{1-c_{\rm eff}(r)}{24}}.
\eeq
The reflection phase $\bar\de^{\rm (TBA)}(P)$ from TBA is now defined 
as the quantization condition, namely,
\beq \label{bd_delta_tba}
\bar\de^{\rm (TBA)} = \pi + 4 \bar{Q} P \ln{\frac{r}{2\pi}}.
\eeq
According to Eq.(\ref{ceffective}),
$\bar\de^{\rm (TBA)}(P)$ should reproduce the Liouville phase $\bar\de(P)$
Eq.(\ref{quantize}) up to exponentially small corrections in $1/P$,
\[
\bar\de(P) = \bar\de^{\rm (TBA)} + {\cal O}(e^{-\frac{\pi}{2P\bar{Q}}}).
\]
In Table 1, we show the first three coefficients of $\bar\de^{\rm (TBA)}$ in
the expansion in powers of $P$ obtained by numerical analysis of the BD model 
and compare with the corresponding $\bar\de^{\rm (LFT)}$ given by 
Eq.~(\ref{bd_delta}). We see an excellent agreement which confirms the
validity of our approach.

\begin{table}
\begin{tabular}{||c||c|c||c|c||c|c||} \hline
\rule[-.4cm]{0cm}{1.cm}B   & $\bar\de^{\rm (TBA)}_1$ 
& $\bar\de^{\rm (LFT)}$ &
$\bar\de^{\rm (TBA)}_3$ & $\bar\de^{\rm (LFT)}$ &
$\bar\de^{\rm (TBA)}_5$ & $\bar\de^{\rm (LFT)}$ \\ \hline
0.3 & --11.2     &  --11.321358 &           &   138.3448 &
 & --2959.790 \\ 
0.4 & --6.823    &  --6.8231877 &   81.   &   82.87330 &
              & --1240.056 \\
0.5 & --4.102542 &  --4.1025435 &   54.95 &   54.96262 &
      --580 & --605.9742 \\
0.6 & --2.3474296 & --2.3474296 &   39.2161 & 39.21607 &
      --327. & --326.6206 \\
0.7 & --1.1951176 & --1.1951176 &   29.8445 & 29.84440 &
      --190. & --190.1987 \\
0.8 & --0.46308101 & --0.46308101 & 24.2903 & 24.29028 &
      --121. & --120.7318 \\
0.9 & --0.05521109 & --0.05521108 & 21.3308 & 21.33073 &
      --87.6 & --87.37743 \\
1.0 &  0.07595095 & 0.07595096 &  20.3996 &  20.39958 &
      --77.6 & --77.42789 \\ \hline
\end{tabular}
\caption{First three coefficients of $\bar\de^{\rm (TBA)}$ in the expansion 
in powers of $P$ obtained by numerical analysis of the BD model in 
comparison with
the corresponding $\bar\de^{\rm (LFT)}$ given by Eq.(\ref{bd_delta}).}
\end{table}

We have also plotted in Fig.1 the scaling function $c_{\rm eff}(r)$ 
obtained from numerical analysis of TBA equations and that from LFT reflection
amplitudes (\ref{ceffective}), or equivalently, the full 
analytic evaluation of TBA equation (\ref{trunceff}) with ${\cal O}(r)$ 
power corrections neglected.
we find that they agree for $r \simeq 0.1$ beyond which ${\cal O}(r)$ 
power correction becomes important.

\begin{figure}
\rotatebox{-90}{\resizebox{!}{15cm}{\scalebox{0.1}{%
{\includegraphics[4cm,2cm][31cm,25cm]{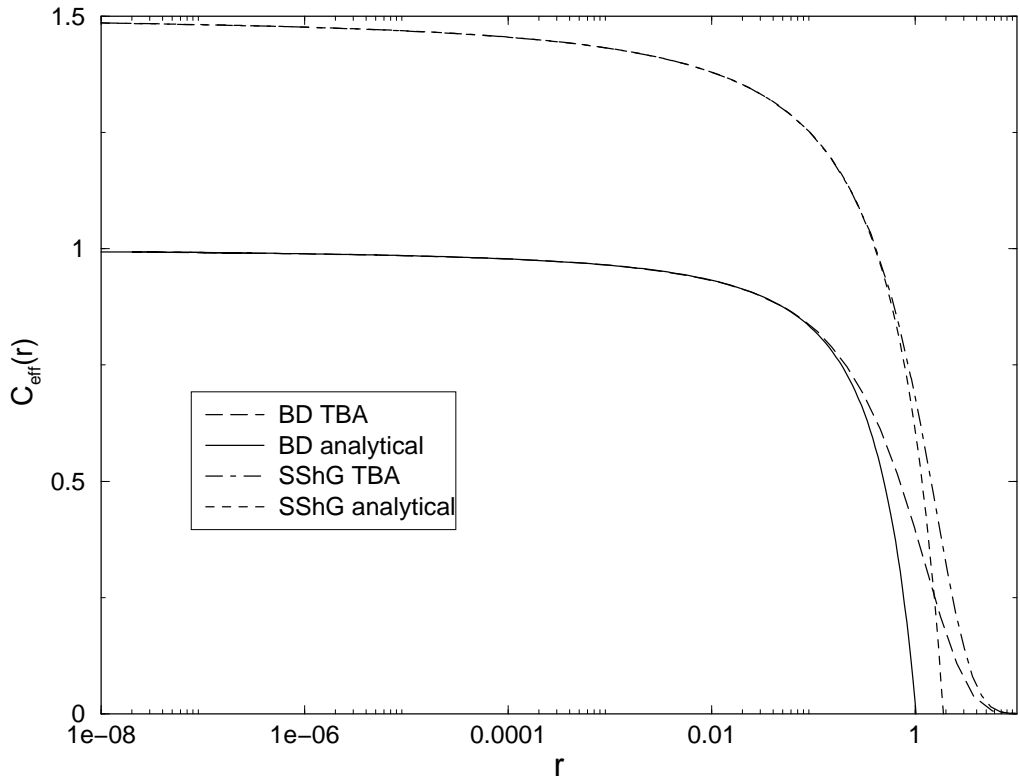}}}}}
\caption{Plot of $c_{\rm eff}$ for the BD and the SShG models. $B = 0.5$ for
the BD model and $B = 0.3$ for the SShG model.}
\end{figure}
We can do the similar analysis for the (NS) sector of the SShG model. 
We define the momentum $P$ in TBA framework as
\[
P = \sqrt{\frac1{12}\left( \frac32 - c_{\rm eff}(r)\right)}\,,
\]
and $\de^{\rm (TBA)}$ as
\[
\de^{\rm (TBA)}(P)=\pi+2QP\ln{r\over{2\pi}}.
\]

\begin{table}
\begin{tabular}{||c|c||c||c|c||c|c||} \hline
\rule[-.4cm]{0cm}{1.cm} B   & $\de^{\rm (TBA)}_1$ & $\de^{\rm (SLFT)}_1$ &
          $\de^{\rm (TBA)}_3$ & $\de^{\rm (SLFT)}_3$ &
          $\de^{\rm (TBA)}_5$ & $\de^{\rm (SLFT)}_5$ \\ \hline
 0.1 &             &--3.2785336087&            &  21.66670467
               &           & --100.7910846 \\
 0.18& 1.0670       &1.06690190762& 7.8        &  7.874361330
               &           & --18.38159324 \\
 0.2 & 1.63253      &1.63252399743& 6.50       &  6.511141559
               &           & --13.28563686 \\
 0.22& 2.095776     &2.09577590830& 5.469      &  5.469890529
               & --9.6     & --9.834738657 \\
 0.24& 2.4799767    &2.47997667482& 4.65801    &  4.658037011
               & --7.41    & --7.424651930 \\
 0.26& 2.801552016  &2.80155201512& 4.014772   &  4.014774388
               & --5.697   & --5.698671053 \\
 0.28& 3.0723945744 &3.07239457437& 3.498766   &  3.498766117
               & --4.437   & --4.437010555 \\
 0.3 & 3.3013090796 &3.30130907959& 3.0811053  &  3.081105362
               & --3.4992  & --3.499327881 \\
 0.32& 3.4949316196 &3.49493161957& 2.7411016  &  2.741101692
               & --2.7932  & --2.793281686 \\
 0.34& 3.6583334039 &3.65833340390& 2.4636642  &  2.463664336
               & --2.2565  & --2.256567975 \\
 0.36& 3.7954280414 &3.79542804138& 2.2376252  &  2.237625215
               & --1.8462  & --1.846269137 \\
 0.38& 3.9092523897 &3.90925238965& 2.0546356  &  2.054635580
               & --1.5323  & --1.532335806 \\
 0.4 & 4.0021635972 &4.00216359715& 1.9084242  &  1.908424295
               & --1.29347 & --1.293489949 \\
 0.42& 4.07597900793&4.07597900793& 1.7942911  &  1.794291077
               & --1.11458 & --1.114593485 \\
 0.44& 4.13207600398&4.13207600398& 1.70875696 &  1.708756972
               & --0.98492 & --0.984931878 \\
 0.46& 4.17146289710&4.17146289710& 1.64932333 &  1.649323338
               & --0.89708 & --0.897087371 \\
 0.48& 4.19482815387&4.19482815387& 1.61430848 &  1.614308492
               & --0.84620 & --0.846206262 \\
 0.5 & 4.20257268596&4.20257268596& 1.60274253 &  1.602742538
               & --0.82954 & --0.829542204 \\ \hline
\end{tabular}
\caption{First three coefficients of $\de^{\rm (TBA)}$ in the expansion 
in powers of $P$ obtained by numerical analysis of the SShG model in comparison 
with the corresponding $\de^{\rm (SLFT)}$ given by Eq.(\ref{ns_phase}).}
\end{table}

Table 2 shows the first three coefficients of $\de^{\rm (TBA)}$ in
the power expansion in $P$ obtained by numerical analysis of the SShG model 
and the corresponding $\de^{\rm (SLFT)}$ given by Eq.~(\ref{ns_phase})
supplemented with the $\mu-m$ relation (\ref{ssgmmu}). Here again we see 
excellent agreements between the numerical and the analytical results. 
This result shows that the method based on the reflection amplitudes
works perfectly in the presence of fermions like the SShG model.
Especially the quantization condition for $P$ holds perfectly even for $b$ 
not small for which fermions interact nontrivially with bosons and the 
validity of the quantization condition may not be entirely clear at the first 
look. Thus our numerical result fully supports the analysis of
TBA equations in sect.~4 and also the identification of undetermined 
constant $\be$ as that coming from SLFT reflection amplitude.
Actually, from the plot of $c_{\rm eff}(r)$ in Fig.\ 1, the approximation 
$c_{\rm eff} = 3/2 - 12 P^2$ is seen to work for larger
region of $r$ than that in the BD case. 
The supression of ${\cal O}(r)$ correction might 
be related with the supersymmetric nature of the model.

\vskip 1cm

\section{Conclusion}

In this paper we have analyzed the scaling functions of 
the ShG, the BD and the SShG models considered as integrable
perturbations of the LFT and super-LFT using 
the reflection amplitudes and conventional TBA analysis.
Our main result is that the new method based on the reflection
amplitudes is not only consistent with the TBA but also 
necessary to completely understand the UV behaviour of the TBA.
As our perturbative computations of the TBA show, there are 
no higher order corrections in the scaling functions except the 
leading correction which includes an unknown constant.
To fix this constant, one should compare the scaling function
obtained by the reflection amplitude, in particular 
the quantization conditions with the condition that the 
$L$ function in the TBA analysis be symmetric.
Indeed, we identified the hidden condition in the TBA which 
are equivalent to the quantization condition.
Our analytic results are fully supported by numerical studies
where we have correctly reconstructed the reflection amplitudes
from the TBA.

In addition, we find that the UV behaviours of the (NS) and (R) sectors 
in the SShG model are quite different. 
The scaling function of the (R) sector 
has only the power law perturbative corrections which vanish
as $b\to 0$.

There are some interesting open problems which can be answered in 
a similar method used in this paper.
First, our analysis of the ShG model, the $A^{(1)}_1$ affine Toda field theory,
can be extended to
those with higher rank by generalizing the one-dimensional quantum 
mechanical quantization condition to the higher-dimensional 
one \cite{AKRY}.

Second, the new method to compute the scaling functions in terms
of the reflection amplitudes can
be extended to give the scaling behviour of the central charge of  
non-integrable models which can be expressed as
perturbations of the LFT and super LFT.
Previously, the scaling functions of the central charges can be
computed only when the models are integrable so that one can use
the TBA methods.

Another problem is to generalize the quantization condition and 
corresponding quantum mechanical interpretation so that the 
scaling function obtained in this way makes sense for the whole range of
the scale $R$. 
To do this, one may need non-perturbative expression for
the power corrections.

So far we have considered only the bulk integrable models.
There has been much progress recently in the integrable models
with boundaries.
The main physical quantity which can be computed using boundary TBA
is the boundary entropy which can also flow as the boundary
scale varies.
Using a similar logic presented in this paper, one may compute the
scaling function of the boundary entropy using the reflection
amplitudes of the boundary correlation functions for the boundary
LFT \cite{FLZZ2}.

We hope our approach made in this paper can provide an alternative
approach to compute the scaling functions in the two dimensional
quantum field theories.

\section*{\bf Acknowledgement}

We thank V. Fateev, A. Fring, and Al. Zamolodchikov for valuable 
discussions. We also thank APCTP, KIAS and CA thanks Freie 
Univ. in Berlin, and Univ. Montpellier II for hospitality.
This work is supported in part by Alexander
von Humboldt foundation and the Grant for the Promotion of Scientific
Research in Ewha Womans University (CA) and Korea Research Foundation
1998-015-D00071 (CR).

\end{document}